\pdfoutput=1

\documentclass[11pt]{article}

\usepackage{naacl2021}

\usepackage{times}
\usepackage{latexsym}
\usepackage{stfloats}
\usepackage[T1]{fontenc}

\usepackage[utf8]{inputenc}

\usepackage{microtype}

\usepackage{graphicx}
\usepackage[misc,geometry]{ifsym}
\usepackage{multirow}
\usepackage{enumitem}
\usepackage{subcaption}
\usepackage{booktabs}
\usepackage{siunitx}

\title{Bootstrapping Large-Scale Fine-Grained Contextual Advertising Classifier from Wikipedia}

\author{Yiping Jin$^{1,2}$, Vishakha Kadam$^1$, Dittaya Wanvarie$^2$ \\
$^1$Knorex, 140 Robinson Road, 14-16 Crown @ Robinson, Singapore\\
$^2$Department of Mathematics \& Computer Science, Chulalongkorn University, Thailand\\
\texttt{\{jinyiping, vishakha.kadam@\}@knorex.com} \\
\texttt{Dittaya.W@chula.ac.th} \\
}

\begin{document}
\maketitle
\begin{abstract}
Contextual advertising provides advertisers with the opportunity to target the context which is most relevant to their ads. The large variety of potential topics makes it very challenging to collect training documents to build a supervised classification model or compose expert-written rules in a rule-based classification system. Besides, in fine-grained classification, different categories often overlap or co-occur, making it harder to classify accurately. 

In this work, we propose \textit{wiki2cat}, a method to tackle large-scaled fine-grained text classification by tapping on the Wikipedia category graph. The categories in the IAB taxonomy are first mapped to category nodes in the graph. Then the label is propagated across the graph to obtain a list of labeled Wikipedia documents to induce text classifiers. The method is ideal for large-scale classification problems since it does not require any manually-labeled document or hand-curated rules or keywords.
The proposed method is benchmarked with various learning-based and keyword-based baselines and yields competitive performance on publicly available datasets and a new dataset containing more than 300 fine-grained categories.
\end{abstract}

\section{Introduction}

Despite the fast advancement of text classification technologies, most text classification models are trained and applied to a relatively small number of categories. Popular benchmark datasets contain from two up to tens of categories, such as SST2 dataset for sentiment classification (2 categories)~\cite{socher2013recursive}, AG news dataset (4 categories)~\cite{zhang2015character} and 20 Newsgroups dataset~\cite{lang1995newsweeder} for topic classification. 

In the meantime, industrial applications often involve fine-grained classification with a large number of categories. For example, Walmart built a hybrid classifier to categorize products into 5000+ product categories~\cite{sun2014chimera}, and Yahoo built a contextual advertising classifier with a taxonomy of around 6000 categories~\cite{broder2007semantic}. Unfortunately, both systems require a huge human effort in composing and maintaining rules and keywords. Readers can neither reproduce their system nor is the system or data available for comparison.

\begin{figure}[!ht]
  \centering \includegraphics[width=200pt]{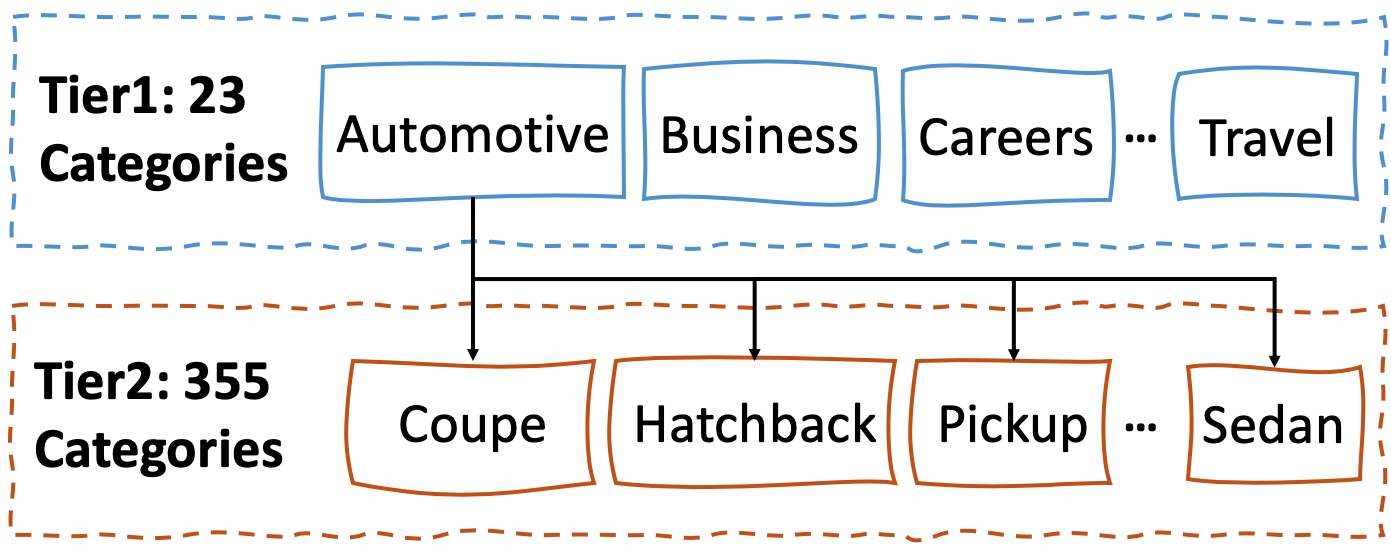}
  \caption{Snippet of IAB Content Categorization Taxonomy.}
  \label{fig:iab-taxonomy}
\end{figure}

In this work, we focus on the application of contextual advertising~\cite{jin2017combining}, which allows advertisers to target the context most relevant to their ads. However, we cannot fully utilize its power unless we can target the page content using fine-grained categories, e.g., ``coup\'e''' vs. ``hatchback'' instead of ``automotive'' vs. ``sport''. This motivates a classification taxonomy with both high coverage and high granularity. The commonly used contextual taxonomy introduced by Interactive Advertising Bureau (IAB) contains 23 coarse-grained categories and 355 fine-grained categories~\footnote{\url{https://www.iab.com/guidelines/taxonomy/}}. Figure~\ref{fig:iab-taxonomy} shows a snippet of the taxonomy.

Large online encyclopedias, such as Wikipedia, contain an updated account of almost all topics. Therefore, we ask an essential question: \textbf{can we bootstrap a text classifier with hundreds of categories from Wikipedia without any manual labeling}? 

We tap on and extend previous work on Wikipedia content analysis~\cite{kittur2009s} to automatically label Wikipedia articles related to each category in our taxonomy by Wikipedia category graph traversal. We then train classification models with the labeled Wikipedia articles. We compare our method with various learning-based and keyword-based baselines and obtain a competitive performance.

\section{Related Work}
\label{sec:related_work}

\subsection{Text Classification Using Knowledge Base}
\label{subsec::large-scale}

Large knowledge bases like Wikipedia or DMOZ content directory cover a wide range of topics. They also have a category hierarchy in either tree or graph structure, which provides a useful resource for building text classification models. Text classification using knowledge bases can be broadly categorized into two main approaches: vector space model and semantic model.

Vector space model aims to learn a category vector by aggregating the descendant pages and perform nearest neighbor search during classification. 
A pruning is usually performed first based on the depth from the root node or the number of child pages to reduce the number of categories. Subsequently, each document forms a document vector, which is aggregated to form the category vector. ~\citet{lee2013semantic} used tf-idf representation of the document, while ~\citet{kim2018incorporating} combined word embeddings and tf-idf representations to obtain a better performance. 

In semantic models, the input document is mapped explicitly to concepts in the knowledge base. The concepts are used either in conjunction with bag-of-words representation~\cite{gabrilovich2006overcoming} or stand-alone~\cite{chang2008importance} to assign categories to the document.

~\citet{gabrilovich2006overcoming} used a feature generator to predict relevant Wikipedia concepts (articles) related to the input document. 
These concepts are orthogonal to the labels in specific text classification tasks and are used to enrich the representation of the input document. Experiments on multiple datasets demonstrated that the additional concepts helped improve the performance. Similarly, ~\citet{zhang2013improving} enriched the document representation with both concepts and categories from Wikipedia. 

~\citet{chang2008importance} proposed Dataless classification that maps both input documents and category names into Wikipedia concepts using Explicit Semantic Analysis~\cite{gabrilovich2007computing}. The idea is similar to Gabrilovich and Markovitch~\shortcite{gabrilovich2006overcoming}, except (1) the input is mapped to a real-valued concept vector instead of a discrete list of related categories, and (2) the category name is mapped into the same semantic space, which removes the need for labeled documents.

Most recently, \citet{chu2020natcat} improved text classification by utilizing naturally labeled documents such as Wikipedia, Stack Exchange subareas, and Reddit subreddits. Instead of training a traditional supervised classifier, they concatenate the category name and the document and train a binary classifier, determining whether the document is related to the category. They benchmarked their proposed method extensively on 11 datasets covering topical and sentiment classification.

Our work is most similar to ~\citet{lee2013semantic}. However, they only evaluated on random-split Wikipedia documents, while we apply the model to a real-world large-scale text classification problem. We also employed a graph traversal algorithm to label the documents instead of labeling all descendant documents.

\subsection{Wikipedia Content Analysis}
\label{subsec::content-analysis}

Some previous work tried to understand the distribution of topics in Wikipedia for data analysis and visualization~\cite{mesgari2015sum}. \citet{kittur2009s} calculated the distance between each page to top-level category nodes. They then assigned the category with the shortest distance to the page. With this approach, they provided the first quantitative analysis of the distribution of topics in Wikipedia. 

\citet{farina2011automatically} extended the method by allowing traversing upward in the category graph and assigning categories proportional to the distance instead of assigning the category with the shortest-path only. 
More recently, ~\citet{bekkerman17} visualized Wikipedia by building a two-level coarse-grained/fine-grained graph representation. 
The edges between categories capture the co-occurrence of categories on the same page. They further pruned edges between categories that rarely appear together. The resulting graph contains 441 largest categories and 4815 edges connecting them. 

\section{Method}
\label{sec:method}

\begin{figure*}
  \centering \includegraphics[width=0.9\textwidth]{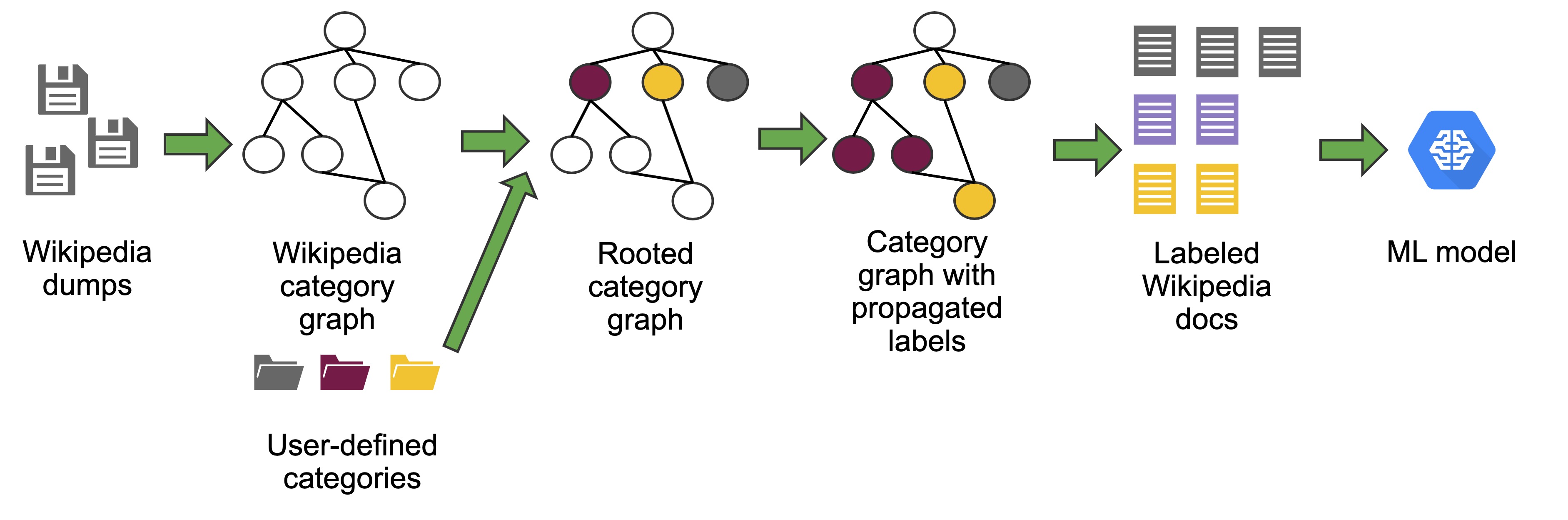}
  \caption{Overview of wiki2cat, a framework to bootstrap large-scale text classifiers from Wikipedia. We first map user-defined categories to category nodes in the Wikipedia category graph. Then, we traverse the category graph to label documents automatically. Lastly, we use the labeled documents to train a supervised classifier.}
  \label{fig:overview}
\end{figure*}

We propose \textit{wiki2cat}, a simple framework using \textbf{Wiki}pedia to bootstrap text \textbf{cat}egorizers. We first map the target taxonomy to corresponding Wikipedia categories (briefed in Section~\ref{subsec:cat_mapping}). We then traverse the Wikipedia category graph to automatically label Wikipedia articles (Section~\ref{subsec:graph_traversal}). Finally, we induce a classifier from the labeled Wikipedia articles (Section~\ref{subsec:Training}).
Figure~\ref{fig:overview} overviews the end-to-end process of building classifiers under the wiki2cat framework.

\subsection{Mapping the Target Taxonomy to Wikipedia Categories}
\label{subsec:cat_mapping}

Wikipedia contains 2 million categories, which is 4 orders of magnitude larger than IAB taxonomy. We index all Wikipedia category names in Apache Lucene~\footnote{\url{https://lucene.apache.org}} and use the IAB category names to query the closest matches. We perform the following: 1) lemmatize the category names in both taxonomies, 2) index both Wikipedia category names and their alternative names from redirect links (e.g., ``A.D.D.'' and ``Attention deficit disorder''), 3) split conjunction category names and query separately (e.g., ``Arts \& Entertainment'' $\rightarrow$ ``Arts'', ``Entertainment''), and 4) capture small spelling variations with string similarity~\footnote{We use Jaro-Winkler string similarity with a threshold of 0.9 to automatically map IAB categories to Wikipedia categories.}.

Out of all 23 coarse-grained and 355 fine-grained categories in IAB taxonomy, 311 categories (82\%) can be mapped trivially. Their category names either match exactly or contain only small variations. E.g., the IAB category ``Pagan/Wiccan'' is matched to three Wikipedia categories ``Paganism'', ``Pagans'', and ``Wiccans''. One author of this paper took roughly 2 hours to curate the remaining 67 categories manually and provided the mapping to Wikipedia categories. Out of the 67 categories, 23 are categories that cannot be matched automatically because the category names look very different, e.g., ``Road-Side Assistance'' and ``Emergency road services''. The rest are categories where the system can find a match, but the string similarity is below the threshold (e.g., correct: ``Chronic Pain'' and ``Chronic Pain Syndromes''; incorrect: ``College Administration'' and ``Court Administration''). We use the curated mapping in subsequent sections.

\subsection{Labeling Wikipedia Articles by Category Graph Traversal}
\label{subsec:graph_traversal}

With the mapping between IAB and Wikipedia categories, we can anchor each IAB category as nodes in the Wikipedia category graph~\footnote{We construct the category graph using the ``subcat'' (subcategory) relation in the Wikipedia dump. The graph contains both category nodes and page nodes. Pages all appear as leaf nodes while category nodes can be either internal or leaf nodes.}, referred to as the \textit{root category nodes}. Our task then becomes to obtain a set of labeled Wikipedia articles by performing graph traversal from the root category nodes. From each root category node, the category graph can be traversed using the breadth-first search algorithm to obtain a list of all descendant categories and pages. 

One may argue that we can take all descendant pages of a Wikipedia category to form the labeled set. However, in Wikipedia page $A$ belongs to category $B$ does not imply a hypernym relation. In fact, some pages have a long list of categories, most of which are at their best remotely related to the main content of the page. E.g., the page ``Truck Stop Women''~\footnote{\url{https://en.wikipedia.org/wiki/Truck\_Stop\_Women}} is a descendant page of the category ``Trucks''. However, it is a 1974 film, and the main content of the page is about the plot and the cast.

\begin{figure*}[t]
  \centering \includegraphics[width=0.8\textwidth]{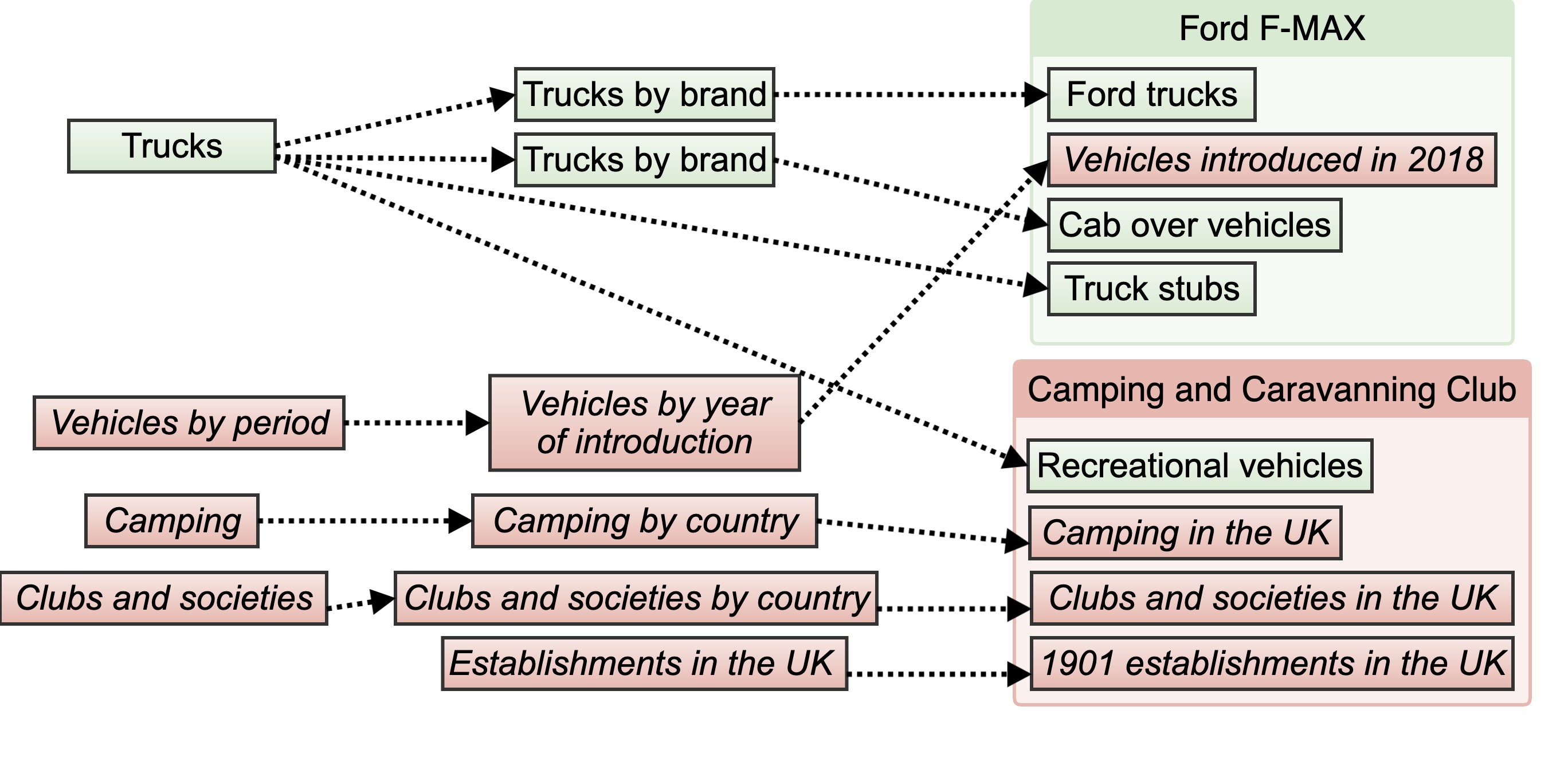}
  \caption{Intuition of the pruning for the category ``Trucks''. The page ``Ford F-Max'' belongs to four categories. Three of which can be traversed from ``Trucks'' and one cannot (marked in red and italic).}
  \label{fig:overlap}
\end{figure*}

We label Wikipedia pages using a competition-based algorithm following ~\citet{kittur2009s} and ~\citet{farina2011automatically}. We treat each category node from which a page can be traversed as a candidate category and evaluate across all candidate categories to determine the final category(s) for the page. 

Firstly, all pages are pruned based on the percentage of their parent categories that can be traversed from the root category. Figure~\ref{fig:overlap} shows two Wikipedia pages with a snippet of their ancestor categories. Both pages have a shortest distance of 2 to the category ``Trucks''. However, the page ``Ford F-Max'' is likely more related to ``Trucks'' than the page ``Camping and Caravanning Club'' because most of its parent categories can be traversed from ``Trucks''. We empirically set the threshold that we will prune a page with respect to a root category if less than 30\% of its parent categories can be traversed from the root category.

While the categories in IAB taxonomy occur in parallel, the corresponding categories in Wikipedia may occur in a hierarchy. For example, the category ``SUVs'' and ``Trucks'' are in parallel in IAB taxonomy but ``SUVs'' is a descendant category of ``Trucks'' in Wikipedia (Trucks \guilsinglright Trucks by type \guilsinglright Light trucks \guilsinglright Sport utility vehicles). While traversing from the root category node, we prune all the branches corresponding to a competing category. 

Pruning alone will not altogether remove the irrelevant content, because the degree of semantic relatedness is not considered. We measure the semantic relatedness between a page and a category based on two factors, namely the shortest path distance and the number of unique paths between them. Previous work depends only on the shortest path distance~\cite{kittur2009s,farina2011automatically}. We observe that if a page is densely connected to a category via many unique paths, it is often an indication of a strong association. We calculate the weight $w$ of a page with respect to a category as follows: 

\begin{equation} \label{eq:category-weight}
w= \displaystyle\sum_{i=0}^{k}\frac{1}{2^{d_i}} 
\end{equation}

where $k$ is the number of unique paths between the page and the category node, and $d_i$ is the distance between the two in the $i$th path. To calculate the final list of categories, the weights for all competing categories are normalized to 1 by summing over each candidate category $j$ and the categories which have a weight higher than 0.3 are returned as the final assigned categories.

\begin{equation} \label{eq:category-weight-normalization}
w_j = \displaystyle\sum_{i=0}^{k_j}\frac{1}{2^{d_{ij}}} / (\displaystyle\sum_{j}\displaystyle\sum_{i=0}^{k_j}\frac{1}{2^{d_{ij}}})
\end{equation}

\begin{figure*}[t]
  \centering \includegraphics[width=0.9\textwidth]{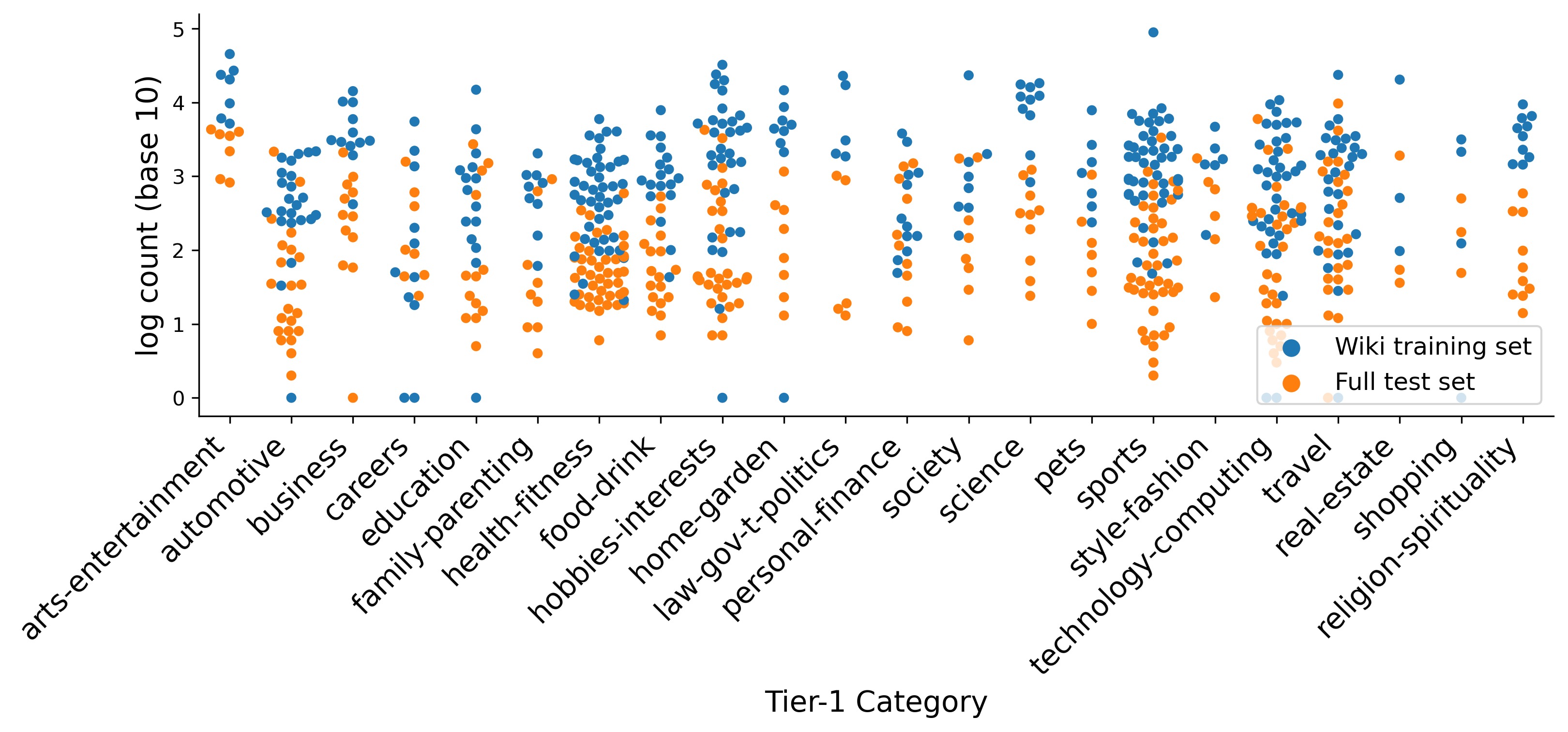}
  \caption{Blue: \# of automatically labeled Wikipedia articles per fine-grained category in log-10 scale. (mean=2.95, std=0.86). Orange: \# of articles per fine-grained category in the full test set in log-10 scale (mean=1.94, std=0.78).}
  \label{fig:labeling_plot}
\end{figure*}

The labeling process labeled in total 1.16 million Wikipedia articles. The blue scattered plot in Figure~\ref{fig:labeling_plot} plots the number of labeled training articles per fine-grained category in log-10 scale. We can see that the majority of the categories have between 100 to 10k articles.

\subsection{Training Contextual Classifiers}
\label{subsec:Training}

The output of the algorithm described in Section~\ref{subsec:graph_traversal} is a set of labeled Wikipedia pages. In theory, we can apply any supervised learning method to induce classifiers from the labeled dataset. The focus of this work is not to introduce a novel model architecture, but to demonstrate the effectiveness of the framework to bootstrap classifiers without manual labeling. We experiment with three simple and representative classification models. The first model is a linear SVM with tf-idf features, which is a competitive baseline for many NLP tasks~\cite{wang2012baselines}. The second model is a centroid classifier, which is commonly used in large-scale text classification~\cite{lee2013semantic}. It averages the tf-idf vectors of all documents belonging to each category and classifies by searching for the nearest category vector. The third model uses BERT~\cite{devlin2018bert} to generate the semantic representation from the text and uses a single-layer feed-forward classification head on top. We freeze the pre-trained BERT model and train only the classification head for efficient training. 

The number of labeled Wikipedia documents for each category is highly imbalanced. Minority categories contain only a handful of pages, while some categories have hundreds of thousands of pages. We perform random over- and downsampling to keep 1k documents for each fine-grained category and 20k documents for each coarse-grained category to form the training set.~\footnote{We use the original dataset without sampling for the centroid classifier since it is not affected by label imbalance.} 

\section{Experiments}
\label{sec:experiment}

\subsection{Evaluation Datasets}
\label{subsec:datasets}

We evaluated our method using three contextual classification datasets. The first two are coarse-grained evaluation datasets published by ~\citet{jin2019learning} covering all IAB tier-1 categories except for ``News'' (totaling 22 categories). The datasets are collected using different methods (news-crawl-v2 dataset (nc-v2) by mapping from news categories; browsing dataset by manual labelling) and contain 2,127 and 1,501 documents separately~\footnote{\url{https://github.com/YipingNUS/nle-supplementary-dataset}}. 

We compiled another dataset for fine-grained classification comprising of documents labeled with one of the IAB tier-2 categories. The full dataset consists of 134k documents and took an effort of multiple person-year to collect. The sources of the dataset are news websites, URLs occurring in the online advertising traffic and URLs crawled with keywords using Google Custom Search~\footnote{\url{https://developers.google.com/custom-search/}}. 

The number of documents per category can be overviewed in Figure~\ref{fig:labeling_plot} (the orange scatter plot). 23 out of 355 IAB tier-2 categories are not included in the dataset because they are too rare and are not present in our data source. So there are in total 332 fine-grained categories in the datasets. Due to company policy, we can publish only a random sample of the dataset with ten documents per category~\footnote{\url{https://github.com/YipingNUS/contextual-eval-dataset}}. We report the performance on both datasets for future work to reproduce our result.  To our best knowledge, this dataset will be the only publicly available dataset for fine-grained contextual classification.

\begin{figure}[!ht]
  \centering \includegraphics[width=0.48\textwidth]{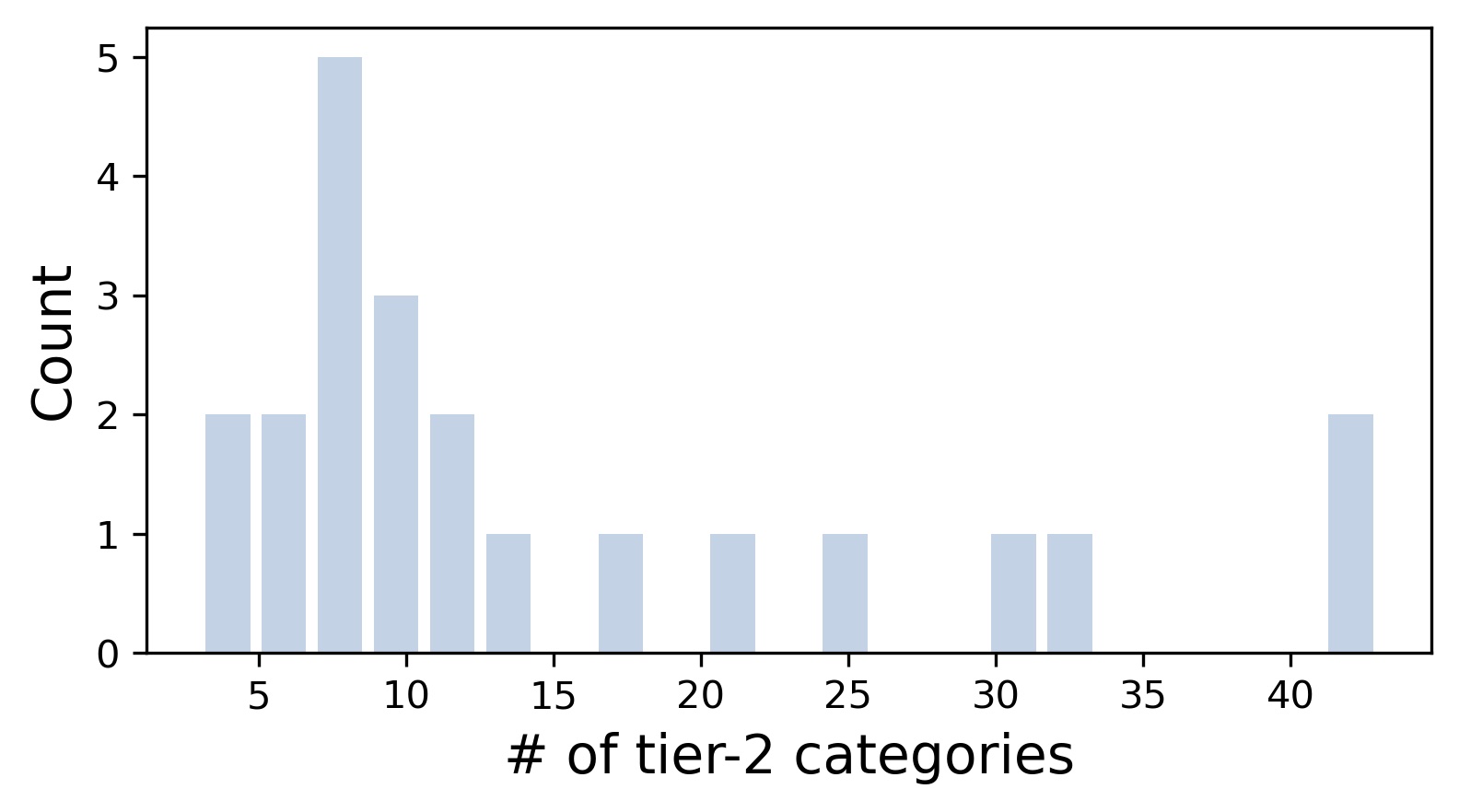}
  \caption{Number of fine-grained categories per coarse-grained category in our fine-grained contextual classification evaluation dataset.}
  \label{fig:t2-hist}
\end{figure}

We focus on classifying among fine-grained categories under the same parent category. Figure~\ref{fig:t2-hist} shows the number of fine-grained categories under each coarse category. While the median number of categories is 10, the classification is challenging because categories are similar to each other.

\subsection{Experimental Settings}
\label{subsec:experiment-setting}

Throughout this paper, we use the Wikipedia dump downloaded on 10 December 2019. After removing hidden categories and list pages, the final category graph contains 14.9 million articles, 1.9 million categories and 37.9 million links. The graph is stored in Neo4J database~\footnote{\url{https://neo4j.com}} and occupies 4.7GB disk space (not including the page content). 

We use the SGD classifier implementation in scikit-learn~\footnote{\url{https://scikit-learn.org}} with default hyperparameters for linear SVM. Words are weighted using tf-idf with a minimum term frequency cutoff of 3. We implement the centroid classifier using TfidfVectorizer in scikit-learn and use numpy to implement the nearest neighbor classification. 

For BERT, we use DistilBERT implementation by HuggingFace~\footnote{\url{ https://huggingface.co/transformers/model\_doc/distilbert.html}}, a model which is both smaller and faster than the original BERT-base model. We use a single hidden layer with 256 units for the feed-forward classification head. The model is implemented in PyTorch and optimized with Adam optimizer with a learning rate of 0.01.

We compare wiki2cat with the following baselines:

    \setlist{nolistsep}
    \begin{itemize}[noitemsep]
    	\item \textbf{Keyword voting} (kw voting): predicts the category whose name occurs most frequently in the input document. If none of the category names is present, the model predicts a random label.
        \item \textbf{Dataless}~\cite{chang2008importance}: maps the input document and the category name into the same semantic space representing Wikipedia concepts using Explicit Semantic Analysis (ESA)~\cite{gabrilovich2007computing}.
        \item \textbf{Doc2vec}~\cite{le2014distributed}: similar to the Dataless model. Instead of using ESA, it uses doc2vec to generate the document and category vector.
        \item \textbf{STM}~\cite{li2018seed}: seed-guided topic model. The state-of-the-art model on coarse-grained contextual classification. Underlying, STM calculates each word's co-occurrence and uses it to ``expand'' the knowledge beyond the given seed words. For coarse-grained classification, STM used hand-curated seed words while STM,S$_{label}$ used category names as seed words. Both were trained by \citet{jin2019learning} on a private in-domain dataset. We also trained STM using our Wikipedia dataset, referred to as STM,D$_{wiki}$. For fine-grained classification, we report only the result of STM,S$_{label}$ since no previously published seed words are available.
    \end{itemize}
    
Keyword voting and Dataless do not require any training document. Both Doc2vec and STM require unlabeled training corpus. We copy the coarse-grained classification result for Doc2vec, STM, and STM,S$_{label}$ from ~\citet{jin2019learning}. For fine-grained classification, we train Doc2vec and STM,S$_{label}$ using the same set of Wikipedia documents as in wiki2cat.

\subsection{Result of Coarse-Grained Contextual Classification}
\label{subsec:experiment-contextual-coarse}
 
We present the performance of various models on nc-v2 and browsing dataset in Table~\ref{tab:coarse-result}. 

\begin{table}[!htbp]
\centering 
\begin{tabular}{p{0.14\textwidth}p{0.05\textwidth}p{0.05\textwidth}p{0.05\textwidth}p{0.05\textwidth}}
\cline{1-5}
\multirow{2}{*}{Model} &\multicolumn{2}{l}{nc-v2} & \multicolumn{2}{l}{browsing}\\ 
&  acc & ma$F_1$& acc$^+$ & ma$F_1$ \\ \cline{1-5}
kw voting & .196 & .180 & .251 & .189 \\
Dataless & .412 & .377 & .536 & .392 \\
Doc2vec & .480 & .461 & .557 & .424 \\
STM & \textbf{.623} & \textbf{.607} & \textbf{.794} & \textbf{.625} \\ 
STM,S$_{label}$ & .332 & .259 & .405 & .340 \\ 
STM,D$_{wiki}$ & .556 & .533 & .780 & .595 \\ 
\cline{1-5}
w2c$_{svm}$ & .563 & .539 & .659 & .523 \\
w2c$_{centroid}$ & .471 & .426 & .675 & .523 \\
w2c$_{bert}$ & .440 & .403 & .621 & .482 \\
\cline{1-5}
\end{tabular}
\caption{Performance of various models on IAB coarse-grained classification datasets. The best performance is highlighted in bold.}
\label{tab:coarse-result}
\end{table}

We can observe that wiki2cat using SVM as the learning algorithm outperformed Dataless and Doc2vec baseline. However, it did not perform as well as STM. The STM model was trained using a list of around 30 carefully chosen keywords for each category. It also used in-domain unlabeled documents during training, which we do not use. ~\citet{jin2019learning} demonstrated that the choice of seed keywords has a significant impact on the model's accuracy. STM,S$_{label}$ is the result of STM using only unigrams in the category name as seed keywords. Despite using the same learning algorithm as STM, its performance was much worse than using hand-picked seed words.


To investigate the contribution of the in-domain unlabeled document to STM's superior performance, we trained an STM model with the manually-curated keywords in ~\citet{jin2019learning} and the Wikipedia dataset we used to train wiki2cat (denoted as STM,D$_{wiki}$). There is a noticeable decrease in performance in STM,D$_{wiki}$ without in-domain unlabeled documents. It underperformed w2c$_{svm}$ on nc-v2 dataset and outperformed it on browsing dataset. 

w2c$_{centroid}$ performed slightly better than w2c$_{svm}$ on the browsing dataset but worse on the nc-v2 dataset. Surprisingly, BERT did not perform as well as the other two much simpler models. We conjecture there are two possible causes. Firstly, BERT has a limitation of sequence length (maximum 512 words). The average sequence length of news-crawl-v2 and browsing datasets are 1,470 and 350 words. Incidentally, there was a more substantial performance gap between BERT and SVM on the news-crawl-v2 dataset. Secondly, our training corpus consists of only Wikipedia articles, while the model was applied to another domain. Therefore, the contextual information that BERT captured may be irrelevant or even counterproductive. We leave a more in-depth analysis to future work and adhere to the SVM and Centroid model hereafter.

\subsection{Impact of Graph Labeling Algorithms}
\label{subsec:experiment-graph-label}

\begin{table}[!htbp]
\centering 
\begin{tabular}{p{0.16\textwidth}p{0.05\textwidth}p{0.05\textwidth}p{0.05\textwidth}p{0.04\textwidth}}
\cline{1-5}
\multirow{2}{*}{Model} &\multicolumn{2}{l}{nc-v2} & \multicolumn{2}{l}{browsing}\\ 
&  acc & ma$F_1$& acc$^+$ & ma$F_1$ \\
\cline{1-5}
w2c & \textbf{.563} & \textbf{.539} & \textbf{.659} & \textbf{.523} \\
\cline{1-5}
w2c$_{child}$ & .325 & .289 & .340 & .322 \\
w2c$_{descendant}$ & .539 & .503 & .607 & .481 \\
w2c$_{min-dist}$ & .533 & .498 & .612 & .489 \\
w2c$_{no-pruning}$ & .488 & .466 & .608 & .491 \\
\cline{1-5}
\end{tabular}
\caption{Performance of the SVM model trained with datasets labeled using different labeling algorithms.}
\label{tab:coarse-result-svm}
\end{table}

We now turn our attention to the impact of different graph labeling algorithms on the final classification accuracy. We compare our graph labeling method introduced in Section~\ref{subsec:graph_traversal} with three methods mentioned in previous work, namely labeling only immediate child pages (child), labeling all descendant pages (descendant), assigning the label with shortest distance (min-dist) as well as another baseline removing the pruning step from our method (no-pruning). We use an SVM model with the same hyperparameters as w2c$_{svm}$. Their performance is shown in Table~\ref{tab:coarse-result-svm}.

Using only the immediate child pages led to poor performance. Firstly, it limited the number of training documents. Some categories have only a dozen of immediate child pages. Secondly, the authors of Wikipedia often prefer to assign pages to specific categories instead of general categories. They assign a page to a general category only when it is ambiguous. Despite previous work in Wikipedia content analysis advocated using shortest distance to assign the topic to articles~\cite{kittur2009s,farina2011automatically}, we did not observe a substantial improvement using shortest distance over using all descendant pages. Our graph labeling method outperformed all baselines, including its modified version without pruning.

\subsection{Result of Fine-Grained Contextual Classification}
\label{subsec:experiment-contextual-fine}

Table~\ref{tab:fine-result} presents the result on fine-grained classification. We notice a performance difference on the full and sample dataset.
However, the relative performance of various models on the two datasets remains consistent. 

\begin{table}[!htbp]
\centering 
\begin{tabular}{lp{1.0cm}p{1.0cm}p{1.0cm}p{0.9cm}}
\cline{1-5}
\multirow{2}{*}{Model} &\multicolumn{2}{l}{Full dataset} & \multicolumn{2}{l}{Sample dataset}\\ 
&  acc & ma$F_1$& acc & ma$F_1$ \\ \cline{1-5}
kw voting & .108 & .018 & .075 & .025 \\
Dataless & .428 & .376 & .477 & .462 \\
Doc2vec & .246 & .152 & .253 & .211 \\
STM,S$_{label}$ & .493 & .370 & .533 & .464 \\ \cline{1-5}
w2c$_{svm}$ & .542$^*$ & \textbf{.464}$^*$ & \textbf{.646}$^*$ & \textbf{.627}$^*$ \\
w2c$_{centroid}$ & \textbf{.548}$^*$ & .451$^*$ & .595$^*$ & .566$^*$ \\
\cline{1-5}
\end{tabular}
\caption{Performance of various models on IAB fine-grained classification datasets. * indicates a statistically significant improvement from baselines with p-value<0.05 using single-sided sample T-test.}
\label{tab:fine-result}
\end{table}

A first observation is that the keyword voting baseline performed very poorly, having 7.5-10.8\% accuracy. It shows that the category name itself is not enough to capture the semantics. E.g., the category ``Travel $>$ South America'' does not match a document about traveling in Rio de Janeiro or Buenos Aires but will falsely match content about ``\textit{South} Korea'' or ``United States of \textit{America}''. 

Dataless and STM outperformed the keyword voting baseline by a large margin. However, wiki2cat is clearly the winner, outperforming these baselines by 5-10\%. It demonstrated that the automatically labeled documents are helpful for the more challenging fine-grained classification task where categories are more semantically similar and harder to be specified with a handful of keywords.


\section{Conclusions and Future Work}
\label{sec:conclusion}

We introduced \textit{wiki2cat}, a simple framework to bootstrap large-scale fine-grained text classifiers from Wikipedia without having to label any document manually. The method was benchmarked on both coarse-grained and fine-grained contextual advertising datasets and achieved competitive performance against various baselines. It performed especially well on fine-grained classification, which both is more challenging and requires more manual labeling in a fully-supervised setting. As an ongoing effort, we are exploring using unlabeled in-domain documents for domain adaptation to achieve better accuracy.

\section*{Acknowledgement}

YJ was supported by the scholarship from `The 100th Anniversary Chulalongkorn
University Fund for Doctoral Scholarship'. We thank anonymous reviewers for their valuable feedback.

\bibliography{anthology,bib}
\bibliographystyle{acl_natbib}




\end{document}